\documentstyle[12pt]{article}

\def\simlt{\mathrel{\lower2.5pt\vbox{\lineskip=0pt\baselineskip=0pt
           \hbox{$<$}\hbox{$\sim$}}}}
\def\simgt{\mathrel{\lower2.5pt\vbox{\lineskip=0pt\baselineskip=0pt
           \hbox{$>$}\hbox{$\sim$}}}}
\def\pl  {{\sl Phys. Lett.}~{\bf B}}

\def\IJMP{{\it Int. J. Mod. Phys. }}
\newcommand{\bea}{\begin{eqnarray}}
\newcommand{\eea}{\end{eqnarray}}
\newcommand{\bean}{\begin{eqnarray*}}
\newcommand{\eean}{\end{eqnarray*}}

\begin{document}
\begin{titlepage}
\vspace*{-1cm}
\hfill{CERN-TH.7169/94}\\
\phantom{bla}
\hfill{hep-th/9402080}\\
\phantom{bla}
\hfill{LPTENS 94/06}
\vskip 1.5cm
\normalsize
\begin{center}
{\Large\bf Construction of Superstrings\\in Wormhole-like
Backgrounds}
\end{center}
\vskip 2cm
\begin{center}
{\bf C. Kounnas}\footnote{On leave from Ecole
Normale
Sup\'erieure, 24 rue Lhomond, 75231 Paris Cedex 05, France.}
\\Theoretical Physics Division, CERN, Geneva, Switzerland\\
\end{center}
\vskip 2cm

\begin{abstract}
\noindent
 We construct a class of superstring solutions in non-trivial
space-time.
The existence of  an $N=4$ world-sheet superconformal symmetry
stabilizes our solutions
under perturbative string loop corrections and implies in target
space some
unbroken space-time supersymmetries.
\end{abstract}
\vspace*{1cm}
\begin{center}
{\it Talk given at the }\\
{\it International Europhysics Conference on High Energy Physics} \\
{\it Marseille, 22-28 July 1993} \\
\end{center}

\vspace*{1cm}
\begin{flushleft} CERN-TH.7169/94 \\
February 1994
\end{flushleft}
\vfill\eject
\end{titlepage}


The study of string propagation in non-trivial gravitational
backgrounds can provide a better understanding of quantum
gravitational phenomena at short distances. Classical string
solutions corresponding to such non-trivial backgrounds can be
obtained by two
different methods. The first makes use of a two-dimensional
$\sigma$-model where the space-time backgrounds correspond to field-
dependent coupling constants. The vanishing of the relevant
$\beta$-functions is identified with the background field equations
of motion in target space [1].
The second approach consists in replacing the free space-time
coordinates by a non-trivial (super)conformal system whose background
interpretation can be seen in the semiclassical limit.
The two methods are useful and complementary. The $\sigma$-model
approach
provides a clear geometric interpretation, but it has the
disadvantage that we can only treat it perturbatively in
$\alpha^{\prime}$. Such a treatment is valid only when all
curvatures and
derivatives of space-time fields are small. In this way, one can
easily obtain approximate solutions, but their possible extension to
exact ones is in general difficult to prove. The conformal field
theory approach takes into account all orders in $\alpha^{\prime}$
automatically and has the main advantage of describing  exact string
vacua.

The background interpretation of a given
exact string solution is a notion that is ill-defined in general
[2],[3].
Indeed, such concepts as space-time dimensionality and topology break
down for
solutions involving highly curved backgrounds, namely when the
metric and/or gauge field curvatures are of the order of the string
scale. When the ``Kaluza-Klein" excitations (quantized ``momentum"
modes) are as massive as the string ``winding"
modes, then the target space interpretation of the string solution
is not as clear, since it is possible to describe the same
string  solution
in terms of non-equivalent backgrounds, which have different topology
and in
some cases different dimensionality as well. This phenomenon is
 intimately
related to the target space duality symmetry among
string solutions [3]-[8], which is well known by now. What in my
opinion
is extremely interesting
is the
equivalence of regular to singular target spaces via string
duality. Once this string phenomenon is well understood, it may
shed some light on the initial singularity problem of classical
cosmological solutions as well as to paradoxes associated with
black holes or other singular objects of classical Einstein gravity.

 In this talk I shall present a special class of exact solutions of
superstrings that are based on some $N=4$
superconformal theories [2]. According to the realization of the
underlying superconformal algebra, our solutions are classified into
several  classes.   More explicitly, we arrange the degrees of
freedom of the ten
supercoordinates in three superconformal systems [2]:
\begin{equation}
\{ \hat c\} = 10 = \{ \hat c = 2 \} + \{ \hat c = 4 \}_1 + \{ \hat c
= 4\}_2~.
\label{ceq}
\end{equation}
The $\hat c = 2$ system is saturated by two free superfields. In one
variation of our solutions, one of the two free superfields is chosen
to be
the time-like supercoordinate and the other to be  one of the nine
space-like ones. In other variations, both
supercoordinates are
Euclidean or even compactified on a one- or two-dimensional torus.
The remaining eight supercoordinates appear in groups of four as
$\{\hat c
= 4\}_1$ and $\{\hat c = 4\}_2$. Both $\{\hat c = 4\}_A$ systems
exhibit
$N = 4$ superconformal symmetry of the Ademollo et al. type [9]. The
non-triviality of our solutions follows from the fact that there
exist realizations of such superconformal
theories, based on spaces with non-trivial geometry and topology,
other
than the
$T^4/Z_2$ orbifold and the $K_3$ manifold.

The first class is characterized by two integer parameters $k_1$,
$k_2$, which are the levels of two $SU(2)$ group manifolds. For
weakly
curved backgrounds (large $k_A$) these solutions can be interpreted
in
terms of a ten-dimensional topologically non-trivial target space
of the form $R^4 \times S^3 \times S^3$. In the special limit
$k_2\rightarrow \infty$ one obtains the semi-wormhole solution of
Callan, Harvey and Strominger [10], based on a six-dimensional flat
background combined with a four-dimensional space
$W_{k_1}^{(4)}\equiv U(1)\times SU(2)_{k_1}$ that describes the
semi-wormhole. The underlying superconformal field theory associated
to $W_{k_1}^{(4)}$ includes a supersymmetric $SU(2)_{k_1}$ WZW model
describing the three coordinates of $S^3$ as well as a non-compact
coordinate with background charge, describing the scale factor of
the sphere. Furthermore, it was known that the {\it five-brane}
background $M^{(6)}\times W_{k_1}^{(4)}$  admits two covariantly
constant
spinors and, therefore, leaves up to two space-time
sypersymmetries unbroken,
consistently with the $N=4$ symmetry of the $W_{k_1}^{(4)}$
superconformal system.
The explicit realization of the desired $N=4$ algebra is derived in
[11], while the target space interpretation as a
four-dimensional semi-wormhole space is given in [10]. In the
context of this interpretation, the 10-d backgrounds
corresponding to the first class of our solutions are products of
topologically non-trivial spaces, $M^{2}\times W_{k_1}^{(4)}\times
W_{k_2}^{(4)}$ ($M^2$ is the flat (1+1) space-time).

A second class of solutions is based on a different realization of
the
$N=4$ superconformal system with $\hat c=4$. Here one replaces the
$W_k^{(4)}$ space by a new $N=4$ system,
$\Delta_k^{(4)} \equiv \bigg\{\bigg (\frac{SU(2)}{U(1)}\bigg)_k
\times \bigg(
\frac{SL(2,R)}{U(1)}\bigg)_{k+4}\bigg\}_{\rm SUSY}$, i.e.
a gauged supersymmetric WZW model, with $\hat c[\Delta^{(4)}_k] = 4$
for any  value of $k$. The choice of the levels $k$ and $k+4$ is
necessary for the existence of an $N=4$ symmetry with $\hat c=4$.
Using $\Delta_k^{(4)}$ or $W_k^{(4)}$ as four-dimensional subspaces,
we can construct non-trivial 10-d solutions, which admit $N=2$ target
space supersymmetries in the heterotic string, or even $N=2+2$ target
space supersymmetries in type-II strings.

Another class of solutions is obtained using the dual space of
$W_k^{(4)}$,
$C^{(4)}_k$ [2],[12],[13]. It turns out that the $C^{(4)}_k$
conformal system with
$\hat c=4$ shares with  $\Delta_k^{(4)}$ and $W_k^{(4)}$ the same
$N=4$ superconformal properties. The explicit realization of  the
$C^{(4)}_k$ space is
given in [2].
{}From the conformal theory viewpoint $C^{(4)}_{k}$  is based on the
supersymmetric gauged WZW model
$C^{(4)}_k \equiv \bigg(\frac{SU(2)}{U(1)}\bigg)_{k}\otimes
U(1)_{R}\otimes U(1)_{Q}$ with a background charge
$Q=\sqrt{\frac{2}{k+2}}$ in  one of the two coordinate currents
($U(1)_{Q}$). The other free  coordinate ($U(1)_{R}$) is compactified
on a torus with radius $R=\sqrt{2k}$.

Having at our disposal non-trivial $N=4$, $\hat c =4$ superconformal
systems,
we can use them as building blocks in order to obtain  new classes of
$exact$
and
$stable$ string solutions in both type
II and
heterotic superstrings. Some typical 10-d
target spaces obtained via the above-mentioned conformal
construction are:
\begin{eqnarray}
{\rm A}) &i) F^{(2)}\otimes W_{k_1}^{(4)} \otimes W_{k_2}^{(4)}
\nonumber\\
&ii) F^{(2)}\otimes F^{(4)} \otimes W_{k}^{(4)}
\nonumber\\
{\rm B}) &i) F^{(2)}\otimes C_{k_1}^{(4)} \otimes C_{k_2}^{(4)}
\nonumber\\
&ii) F^{(2)}\otimes F^{(4)} \otimes C_{k}^{(4)}
\nonumber\\
{\rm C}) &i) F^{(2)}\otimes C_{k_1}^{(4)} \otimes W_{k_2}^{(4)}
\nonumber\\
&ii) F^{(2)}\otimes C_{k_1}^{(4)} \otimes \Delta_{k_2}^{(4)}
\nonumber\\
&iii) F^{(2)}\otimes \Delta_{k_1}^{(4)} \otimes W_{k_2}^{(4)}
\nonumber\\
{\rm D}) &i) F^{(2)}\otimes \Delta_{k_1}^{(4)}
\otimes\Delta_{k_2}^{(4)}
\nonumber\\
&ii) F^{(2)}\otimes F^{(4)} \otimes \Delta_k^{(4)}~.
\label{cons}
\end{eqnarray}
In the above expressions, $F^{(4)}$ stands for a 4-d
flat
space, compact or non-compact, as well as  for the
$T^4/Z_2$ orbifold; $F^{(2)}$ denotes a two-dimensional flat
space, compact or non-compact, with Lorentzian or Euclidean
signature.
Note that the Euclidean version of subclasses A)ii), B)ii), D)ii)
(i.e.
when
$F^{(2)}\otimes F^{(4)}$ is a compact six-dimensional flat space) can
be identified with three different kinds of 4-d
gravitational and/or dilatonic instanton solutions. In this
interpretation, the subspace described by the last factor denotes
the Euclidean
version of our (4-d) space-time.

The type A) constructions based on $W^{(4)}$ conformal theories
describe, from the target space point of view, stable solutions of
4-d gauged supergravities [14], which leave some of the
space-time supersymmetries unbroken. In fact, consider the 10-d
heterotic or
type-II superstring compactified on a product of two
three-dimensional spheres. The corresponding superconformal field
theory is then given by a supersymmetric WZW model based on a ${\bf
K}^{(6)}\equiv SU(2)_{k_1}\otimes SU(2)_{k_2}$ group manifold, where
the affine levels $k_A$ define the radii of the spheres $r_A$,
$k_A
=r_A^2$. In contrast to the toroidal compactification ($T^{(6)}\equiv
U(1)^6$) where the six graviphotons are Abelian, in ${\bf K}^{(6)}$
compactification they become non-Abelian. As expected from field
theory Kaluza-Klein arguments, in the large radius limit the
resulting
effective theory is an $SU(2)_{k_1}\otimes SU(2)_{k_2}$ gauged
supergravity [14]. This can be easily shown in the 2d $\sigma$-model
approach by means of the $\alpha'$-expansion.

The connection with gauged supergravities is very important, because
it allows us to derive the 4-d
effective supergravity action, up to two space-time derivatives,
which is induced by the ${\bf K}^{(6)}$ compactification using
4-d supergravity arguments.

The type A), B) and C) constructions based on $W_{k_A}^{(4)}$,
and
$C_{k_A}^{(4)}$ superconformal systems
are strongly connected to the non-critical superstrings in the
so-called strong coupling regime ($1\leq \hat {c}_{matter} \leq 9$).
 In
fact, the Liouville superfield of non-critical strings can be
identified with the supercoordinate of the above spaces, which has a
non-zero background charge. The central charge of the Liouville
supercoordinate can be easily determined [2],[14],
\begin{equation}
{\hat c}_L=1+2(Q_1^2+Q_2^2)= 1 + 4\left(\frac{1}{k_1+2}
+\frac{1}{k_2+2}\right)\
,
\label{cliou}
\end{equation}
where we have used the relation among the levels $k_A$ and the
background charges $Q_A$, $Q_A^2=2/(k_A+2)$. This relation follows
from the $N=4$ superconformal
symmetry in both $W$ and $C$ systems. The remaining matter part
consists of tensor products of unitary $N=1$ superconformal theories
based on $SU(2)_{k_A}$ WZW, $[SU(2)/U(1)]_{k_A}$ KS cosets, with
$U(1)$ factors. The matter central charge is always given by
\begin{equation}
{\hat c}_M=9 - 4\left(\frac{1}{k_1+2} +\frac{1}{k_2+2}\right)\ ,
\label{cmat}
\end{equation}
and it varies in the region $5\leq{\hat c}_M\leq 9$. Thus, our
explicit constructions show the existence of super-Liouville theories
coupled to $N=1$ superconformal unitary matter systems in the strong
coupling regime. The problematic complex conformal weights, usually
present in this regime, are projected out by the $N=4$ induced
generalized
GSO projection. This projection phenomenon is similar to
the one observed in ref.~[15] in the case of the $N=2$ globally
defined superconformal symmetry.

 In ref.~[2] the reader can find more details about the $N=4$
realizations for the $\hat{c}= 4$ superconformal building blocks.
Here I will present the
target-space  metric together with the
dilaton ($\Phi$) and torsion field strength ($H_{ijk}$) corresponding
to the  $W^{(4)}_{k}$, $C^{(4)}_{k}$, $\Delta^{(4)}_{k}$
4-d subspaces in the large $k$ limit:

\begin{eqnarray}
ds^{2}\left[W_k^{(4)}\right] & = & k
\frac{dzd\bar{z}+dwd\bar{w}}{z\bar{z}+w\bar{w}}
\nonumber \\
-2\Phi & = & \log(z\bar{z}+w\bar{w})+{\rm const.}
\nonumber \\
H_{ijk} & = & e^{+2\Phi}\epsilon_{ijk}~{^{\ell}\partial_{\ell}}\Phi
\end{eqnarray}

\begin{eqnarray}
ds^2\left[C_k^{(4)}\right] & = & k\frac{dzd\bar z}{1-z\bar z } +
k~dwd\bar w
\nonumber \\
-2\Phi & = & w+\bar w +log(1-z\bar z)+{\rm const.}
\end{eqnarray}

It is interesting to note that the $C_k^{(4)}$ space can be obtained
by
performing
a (supersymmetric) duality transformation on $W^{(4)}_{k}$
[12],[13].
In $C_k^{(4)}$ the torsion  is zero, while  it is non-trivial
in  $W^{(4)}_{k}$.
The  metric of $C_k^{(4)}$ is singular, while that
of $W^{(4)}_{k}$ is regular. The question about the relevance of
the
singularity at the stringy level is still an open
question.

 There are two non-equivalent 4-d spaces associated to the
$\Delta_k^{(4)}$ superconformal system that correspond to gauging
either the axial or vector $U(1)$ in the $SL(2,R)/U(1)$ gauged WZW
model.

\begin{eqnarray}
ds^{2}\left[\Delta_k^{(4)}\right]_{\epsilon} =
{}~~~k~\frac{dzd\bar z}{1-z\bar z}
&+& k'~\frac{dwd\bar w}{w\bar w+ \epsilon}~
\nonumber\\
-2\Phi~~ =~\log (1-z\bar z ) &+& \log (w\bar w + \epsilon) + {\rm
const.}~~~~~
\end{eqnarray}

Here $\epsilon=1$ corresponds to the axial gauging and $\epsilon=-1$
to
the
vector one. In the axial case the metric is regular while in the
vector case it is singular. The two versions of $\Delta_k^{(4)}$
spaces
($\epsilon=\pm 1$) are dual to each other. Here also the relevance of
the
singularity is not obvious, but an intriguing issue to understand.

 We hope that our explicit construction of a family of
consistent and
stable solutions will give a better understanding of some fundamental
string properties, especially in the case of strongly curved
backgrounds
(small $k_A$), where the notion of  space-time dimensionality and
topology breaks down.

\vskip 1cm
{\bf \noindent Acknowledgements}

Research supported in part by  the EEC contracts
SC1$^{*}$-0394C and SC1$^{*}$-CT92-0789
\vfill\eject

\end{document}